\documentclass[10pt,conference]{IEEEtran}
\usepackage{graphicx} 
\usepackage{enumitem}
\usepackage{calc}
\usepackage{multicol} 
\usepackage{listings}
\usepackage{adjustbox}

\usepackage{soul}
\usepackage{tabularx}
\usepackage[noend]{algpseudocode}
\usepackage{booktabs}
\usepackage{multirow}
\usepackage{graphicx}
\usepackage{textcomp}
\usepackage{tikz}
\usepackage{xcolor}
\usepackage{enumitem}
\usepackage{caption}
\usepackage[linesnumbered,ruled,lined]{algorithm2e}
\usepackage{alltt}
\usepackage{multirow}
\usepackage{color, soul}
\usepackage{textcomp}
\usepackage{framed}
\usepackage{hhline}
\usepackage{subcaption}
\usepackage[breakable]{tcolorbox}
\usepackage{float}
\usepackage{array}
\usepackage{flushend}
\usepackage{makecell}

\usepackage{amsmath}

\usepackage{balance}
\usepackage{centernot}
\usepackage[T1]{fontenc}
\usepackage[flushleft]{threeparttable}
\usepackage{parcolumns}
\usepackage{url} 
\usepackage{amsthm}
\newtheorem{definition}{Definition}
\usepackage{stmaryrd}
\usepackage[colorinlistoftodos,textwidth=15mm,textsize=tiny,obeyFinal]{todonotes}
\usepackage{lstautogobble}

\usepackage{fancyvrb}
\usepackage{tcolorbox}
\usepackage{hyperref}
\usepackage{diagbox}

\definecolor{azure(colorwheel)}{rgb}{0.0, 0.5, 1.0}

\newtcbox{\mybox}[1][breakable]{on line, enlarge top by=10pt, enlarge bottom by=10pt,
     boxsep=8pt, boxrule=2pt, size=small, arc=1mm}

\definecolor{grey}{rgb}{0.7,0.7,0.7}

\newcommand{\lstbg}[3][0pt]{{\fboxsep#1\colorbox{#2}{\strut #3}}}
\lstdefinelanguage{diff}{
  basicstyle=\ttfamily\scriptsize,,
  morecomment=[f][\lstbg{red!20}]-,
  morecomment=[f][\lstbg{green!20}]+,
  morecomment=[f][\lstbg{yellow!20}]++,
  morecomment=[f][\textit]{@@},
  texcl=false
}

\definecolor{todocolor}{rgb}{0.9,0.1,0.1}
\definecolor{indiagreen}{rgb}{0.07, 0.53, 0.03}
\definecolor{hycolor}{rgb}{0.7,0.7,0.3}
\definecolor{darkbrown}{rgb}{0.4, 0.26, 0.13}

\tikzstyle{highlighter} = [
  yellow,
  line width = \baselineskip,
]

\newcounter{highlight}[page]

\AtBeginShipout{\AtBeginShipoutUpperLeft{\ifthenelse{\value{highlight} > 0}{\tikz[remember picture, overlay]{\foreach \stroke in {1,...,\arabic{highlight}} \draw[highlighter] (begin highlight \stroke) -- (end highlight \stroke);}}{}}}

\newcommand{\ignore}[1]{}
\definecolor{main-color}{rgb}{0.6627, 0.7176, 0.7764}
\definecolor{string-color}{rgb}{0.3333, 0.5254, 0.345}
\definecolor{key-color}{rgb}{0.8, 0.47, 0.196}
\lstdefinestyle{mystyle} {
    language = Java,
    basicstyle = {\ttfamily \color{main-color}},
    stringstyle = {\color{string-color}},
    keywordstyle = {\color{key-color}},
    keywordstyle = [2]{\color{lime}},
    keywordstyle = [3]{\color{yellow}},
    keywordstyle = [4]{\color{teal}},
    morekeywords = [3]{<<, >>},
    morekeywords = [4]{++},
    basicstyle=\ttfamily\scriptsize,
    commentstyle=\color{blue}\ttfamily,
    morecomment=[f][\lstbg{red!20}]-,
    morecomment=[f][\lstbg{green!20}]+,
    morecomment=[f][\lstbg{yellow!20}]++,
    morecomment=[f][\lstbg{yellow!20}]--,
    morecomment=[f][\textit]{@@},
    breaklines=true,
    texcl=false
}
\newcommand{\toolname}{\textsc{RETester}\xspace}

\newcommand{\rebs}{\textsc{REBs}\xspace}

\newcommand{\eclipse}{\textsc{Eclipse}\xspace}
\newcommand{\idea}{\textsc{IntelliJ IDEA}\xspace}

\newcommand{\jdt}{\textsc{Java Development Tools (JDT)}\xspace}
\newcommand{\saferefactor}{\textsc{SAFEREFACTOR}\xspace}
\newcommand{\astgen}{\textsc{ASTGen}\xspace}

\newcommand{\foundIssueNumber}{18}

\newcommand{\totalconfirmedIssueNumber}{seven}
\newcommand{\submittedIssueFixedNumber}{three}

\newcommand{\issueNumberofEclipse}{245}
\newcommand{\issueNumberofIDEA}{213}

\newcommand{\totalIssueNumber}{458}
\newcommand{\compilableIssueNumber}{167}
\newcommand{\uncompilableIssueNumber}{291}

\newcommand{\withTemplateIssueNumber}{15}
\newcommand{\withTemplateEclipseIssueNumber}{13}

\newcommand{\withoutTemplateIssueNumber}{six}
\newcommand{\overlapIssueNumber}{three}
\newcommand{\withoutTemplateNotOverlapIssueNumber}{three}
\newcommand{\withoutTemplateOverlapIssueNumberRatio}{50\%}
\newcommand{\withTemplateNotOverlapIssueNumber}{12}

\newcommand{\eclipseIssueNumber}{18}
\newcommand{\ideaIssueNumber}{three}

\newcommand{\eclipsenotoverlapIssueNumber}{15}

\newcommand{\pullupIssueNumber}{11}

\begin{document}

\title{Testing Refactoring Engine via Historical Bug Report driven LLM}


\author{
\IEEEauthorblockN{Haibo Wang, Zhuolin Xu, Shin Hwei Tan\textsuperscript{*}}
\IEEEauthorblockA{\textit{Department of Computer Science \& Software Engineering}, \textit{Concordia University}, Montreal, Canada \\
haibo.wang@mail.concordia.ca, zhuolin.xu@mail.concordia.ca, shinhwei.tan@concordia.ca}
}

\maketitle

\begin{abstract}
Refactoring is the process of restructuring existing code without changing its external behavior while improving its internal structure. Refactoring engines are integral components of modern Integrated Development Environments (IDEs) and can automate or semi-automate this process to enhance code readability, reduce complexity, and improve the maintainability of software products. Similar to traditional software systems such as compilers, refactoring engines may also contain bugs that can lead to unexpected behaviors. In this paper, we propose a novel approach called \toolname, a LLM-based framework for automated refactoring engine testing. Specifically, by using input program structure templates extracted from historical bug reports and input program characteristics that are error-prone, we design chain-of-thought (CoT) prompts to perform refactoring-preserving transformations. The generated variants are then tested on the latest version of refactoring engines using differential testing. We evaluate \toolname on two most popular modern refactoring engines (i.e., \eclipse, and \idea). It successfully revealed \foundIssueNumber{} new bugs in the latest version of those refactoring engines. By the time we submit our paper, \totalconfirmedIssueNumber{} of them were confirmed by their developers, and \submittedIssueFixedNumber{} were fixed.

\end{abstract}


\section{Introduction}
\label{sec:introduction}

Refactoring is defined as the process of changing a software system in such a way that it does not alter the external behavior of the software, yet improves its internal structure ~\cite{becker1999refactoring}. Refactoring has been well-studied as a way to improve software quality~\cite{du2004refactoring,kim2014empirical} as well as an effective way to facilitate software maintenance and evolution~\cite{kula2018empirical,wahler2016improving}. During software development, developers might perform refactoring manually, which is error-prone and time-consuming, or with the help of tools that automate or semi-automate the activities related to the refactoring process~\cite{mens2004survey, tsantalis2018ten}. Refactoring automation tools like the built-in refactoring engines in IDEs (e.g., \eclipse~\cite{Eclipse}, and \idea~\cite{IDEA}) have been widely used to facilitate software maintenance.

Despite the active development of refactoring engines, they can be buggy. These bugs can silently change the program behaviors, produce uncompilable programs, or induce inconsistencies in the modified code~\cite{wang2024empirical}. 
To ensure the reliability of refactoring engines, several techniques have been proposed to identify Refactoring Engine Bugs (\rebs) via input programs generation, including (1) template-based input program generation technique~\cite{daniel2007automated} that relies on manually crafted imperative generators, and (2) \saferefactor-based tools~\cite{soares2009saferefactor,mongiovi2016scaling,soares2010making,soares2012automated,soares2011identifying,soares2009generating,gligoric2013systematic} that relies that random test generation.  However, there are several limitations in these tools that hinder their effectiveness in identifying REBs. 
Firstly, template-based input program generation techniques like \astgen~\cite{daniel2007automated} usually require developers to manually write a set of predefined templates for generating input programs. However, developers may not have sufficient knowledge nor insights about the types of program structures that are more likely to trigger bugs. Moreover, modern refactoring engines in IDE like \idea support a variety of refactoring types which makes manually designing templates labor-intensive and impractical. Secondly, \saferefactor-based tools~\cite{soares2009saferefactor,mongiovi2016scaling,soares2010making,soares2012automated,soares2011identifying,soares2009generating} rely on automatic test generation tools 
to identify refactoring bugs without considering the bug-triggering ability of the input programs (e.g., input programs contain lambda expression or anonymous class are more error-prone during refactoring), thus leading to low effectiveness. 
    
In recent years, large language model (LLM) trained on programming languages as well as natural languages has shown a great potential to support various software engineering tasks like program repair, code summarization, and code review automation. Prior studies~\cite{shirafuji2023refactoring,choi2024iterative,alomar2024refactor,pomian2024assist} have explored the use of LLM for refactoring activities. However, the feasibility of using LLM to improve the robustness and reliability of the refactoring engines has been under-explored. Given the importance of identifying \rebs and the limitations of
existing approaches, in this paper, we propose a testing approach for refactoring engines called \toolname that leverages historical bug reports together with the bug-triggering input program characteristics, which can be generalizable to support diverse refactoring types. In particular, \toolname leverages historical bug reports, to extract refactoring information. Then, \toolname utilizes LLM combined with prompts engineering to perform Refactoring-preserving Transformations (RPTs). Specifically, by using the input program characteristics that are more error-prone as mutation rules and input program structure templates, we design chain-of-thought (CoT) prompts to perform refactoring-preserving transformations. The generated variants are then tested on the latest version of refactoring engines by applying the same refactoring as in the seed input program. Finally, we use differential testing to find any inconsistencies, and manually inspect each inconsistency before submitting issues to the refactoring engine developers.

Our proposed workflow can benefit refactoring engine testing from two perspectives: First, we leverage the power of LLM to generate diverse bug-triggering input programs based on historical bug reports, which can serve as the first step towards LLM-based testing for refactoring engines. Second, by extracting diverse historical bug-triggering input program structures, template-based techniques (e.g., ASTGen~\cite{daniel2007automated} for refactoring engine testing, JAttack~\cite{zang2022compiler} and LeJit~\cite{zang2024java} for compiler testing) 
can take these error-prone templates as references for designing their templates, thus improving their effectiveness. In summary, this paper makes the following contributions:
\begin{itemize}[leftmargin=*,labelindent=7pt,nosep]
    \item We mined and propose a new dataset that contains human-labeled and high-quality historical bug reports, together with \compilableIssueNumber{} compilable Java programs extracted from these bug reports. Programs in our dataset can be used as the seed inputs for future research on testing software systems that takes Java programs as input (e.g., refactoring engines and compilers). 
    \item We propose a novel LLM-based refactoring engine bug detection approach that (1) automatically mines historical bug reports and extracts high-quality bug-triggering input programs via LLM, (2) performs refactoring-preserving mutations of input programs to obtain similar bug-triggering programs, and (3) leverages prompt template that extracts refactoring information (e.g., refactoring types, and program locations to perform refactoring) from historical bug reports. To the best of our knowledge, we conduct the first systematic study that applies LLM for testing refactoring engine bugs. We open-source our data to facilitate future research in refactoring engine testing~\cite{anonymousrepolink}. 
    \item We conduct experiments on two popular refactoring engines (i.e., \eclipse, and \idea). As a result, we have found \foundIssueNumber{} new bugs in the latest version. By the time we submit our paper, \totalconfirmedIssueNumber{} bugs have been confirmed by their developers, \submittedIssueFixedNumber{} have been fixed.
\end{itemize}

\section{Background}
\label{sec:background}

\subsection{Refactoring Engine}

\begin{figure}[h]
    \centering
    \vspace{-3pt}    \includegraphics[width=0.5\textwidth]{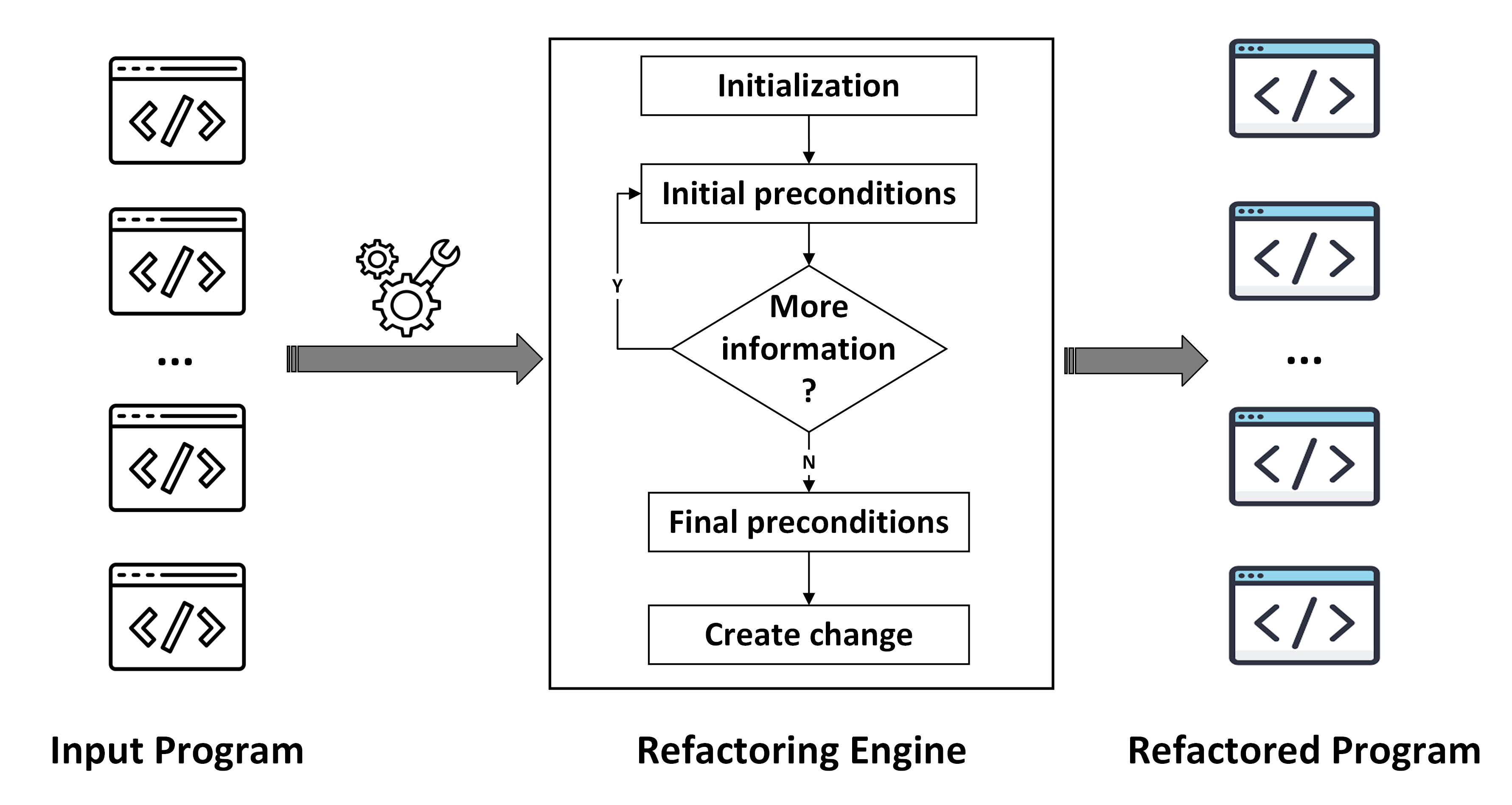}
    \vspace{-6pt}
     \caption{The general workflow of a refactoring engine.}
    \label{refactoringworkflow}
\end{figure}

Refactoring engines, critical components of Integrated Development Environments (IDEs) like \eclipse, and \idea, could automate code restructuring and optimization. Despite varied details and integrations, their workflows share core similarities. The general workflow, depicted in Figure~\ref{refactoringworkflow}, starts with an input program and configuration settings for a specific refactoring and results in the refactored program. The process includes: (1) Initialization, where the engine prepares for refactoring by setting up context and identifying target elements; (2) Initial Conditions Checking, which verifies prerequisites and potential conflicts to ensure refactoring can proceed—if not, errors or warnings are issued; (3) Additional Information Gathering, where the engine collects further inputs or configuration details necessary for the refactoring; (4) Final Conditions Checking, which analyzes the proposed changes, checking for conflicts and verifying if the refactoring preserves behavior and maintains quality—if issues are found, warnings or errors are generated; and finally, (5) Create Change, where the actual code modifications are computed and implemented, including updating declarations, method invocations, associated documentation, and etc. Each stage is designed to ensure that refactoring is executed correctly, maintaining behavior preservation and code syntax.

\subsection{Refactoring Engine Testing}

Several techniques have been proposed to identify refactoring engine bugs~\cite{daniel2007automated,soares2009saferefactor,mongiovi2016scaling,soares2010making,gligoric2013systematic}. Specifically, Daniel et al.~\cite{daniel2007automated} proposed ASTGen, a template-based Java input program generation tool for refactoring engines testing. Soares et al.~\cite{soares2009saferefactor,mongiovi2016scaling,soares2010making} proposed a series of works based on \saferefactor, which relies on random tests generation technique, Randoop, to test the behavior-preserving of refactoring. Gligoric et al.~\cite{gligoric2013systematic} tested Eclipse Java (JDT) and C (CDT) refactoring engines in an end-to-end approach on real software projects. Given a set of projects, they randomly apply refactoring in some program elements and collect failures using differential testing. Eric et al.~\cite{lacker2021statistical} analyzed the existing period, fixing ratio, and duplication percentage of bugs in \eclipse \jdt. Existing works share the following limitations: (1) they are unaware of the bug-triggering ability of the input programs, thus producing or testing on input programs that are less error-prone. (2) manually designing templates for diverse refactoring types and scenarios is time-consuming and labor-intensive, which hinders the scalability. Different from previous works, our work generate refactoring-preserving variant input programs by incorporating the error-prone input program characteristics together with the historical bug-triggering code information. By incorporating the error-prone input program characteristics with the historical bug-triggering input program templates through the refactoring-preserving transformation (Definition~\ref{def:rptdefinition}), new bugs are revealed in the newest version of refactoring engines. Besides, \toolname could tolerant diverse refactoring types without the need of manual designing templates, making it scalable, flexible, and generalizable. To our knowledge, this is the first work demonstrating that LLM can easily perform history-driven testing to the challenging domain of refactoring engine.

\subsection{Motivation Example}


\begin{figure}[h]
    \centering
    \includegraphics[width=0.5\textwidth]{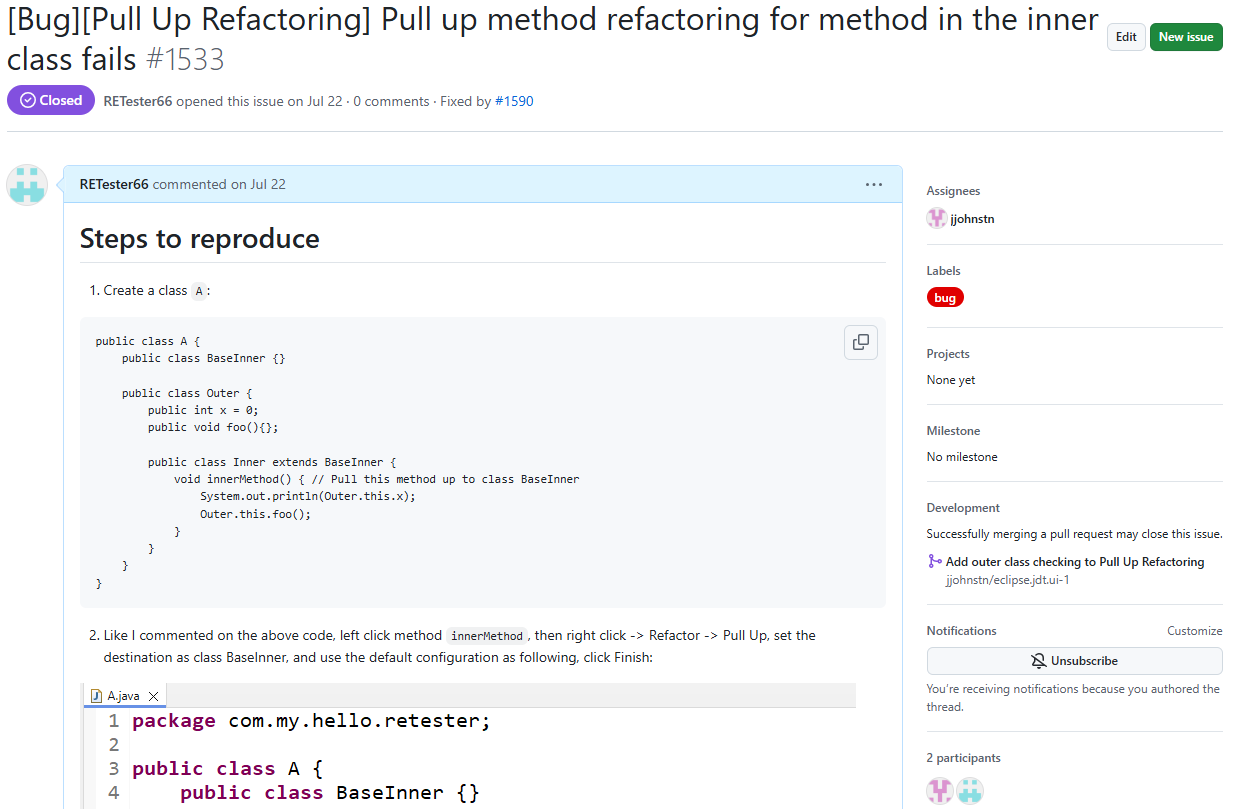}
     \caption{Screenshot of an \eclipse historical bug report~\cite{pullUpRefactoringIssue1533}.}
    \label{bugreportscreenshot}
\end{figure}

\begin{figure}
\centering

\noindent\begin{minipage}{0.49\textwidth}
  \begin{lstlisting}[frame=lines]{Name}
public class A {
    public class BaseInner {}
    public class Outer {
        public int x = 0;
        public void foo(){};
        public class Inner extends BaseInner {
            void innerMethod() { // Pull up to BaseInner
                System.out.println(Outer.this.x);
                Outer.this.foo();
            }}}}
  \end{lstlisting}
\end{minipage}\hfill
\begin{minipage}{.49\textwidth}
  \begin{lstlisting}[frame=lines]{Name}
public class OuterClass {
    public class BaseTargetClass {}
    public class OriginalClass {
        public DataType memberVariable;
        public void memberMethod();
        public class NestedOriginalClass extends BaseTargetClass {
            void methodToBePulledUp() {
            // Method logic that accesses OriginalClass's context
            }}}}
  \end{lstlisting}
\end{minipage}\hfill
\begin{minipage}{.49\textwidth}
  \begin{lstlisting}[frame=lines]{Name}
public class A {
    public class BaseTargetClass {}
    public class OriginalClass {
        public int data = 20;
        public void memberMethod() {}
        public class NestedOriginalClass extends BaseTargetClass {
            void setup() {
                new BaseTargetClass() {
                    void methodToBePulledUp() {
                        System.out.println("Anonymous Class Method: " + data);
                    }
};}}}}
\end{lstlisting}
\end{minipage}
\caption{A bug-triggering input program extracted from historical bug reports (top), extracted template by LLM (middle), and one of its refactoring-preserving variants after applying the Java anonymous class transformation (bottom).} 
\label{inputprogramexample}
\end{figure}

Figure~\ref{bugreportscreenshot} shows the screenshot of one refactoring engine historical bug report from \eclipse~\cite{pullUpRefactoringIssue1533}. As stated in the bug report, refactoring engine of \eclipse fails to resolve the method in a inner class referring to an outer class field when performing the pull up method refactoring, thus producing an uncompilable refactored program. Figure~\ref{inputprogramexample} lists the Java input program (top) extracted from the above historical bug report. Specifically, when applying pull up method refactoring for the method ``innerMethod()'', the refactoring could be successfully performed without any warning message or exception, however, \eclipse would produce an refactored program contains syntax error, thus making the original program or project uncompilable. \eclipse developers have fixed this issue by adding a more reliable dependency analysis to resolve the method in inner class when performing pull up method refactoring~\cite{issuefix1590}.

To challenge refactoring engines under more diverse input programs, our tool generates input programs by incorporating the error-prone input program characteristics and historical bug-triggering input programs from bug reports, which successfully reveals \foundIssueNumber{} new bugs in the latest version of refactoring engines. Specifically, we first construct a dataset containing historical refactoring engine bug reports by mining from refactoring engine bug-tracking systems using keywords (i.e., ``refactoring'' and ``refactor''). To remove the irrelevant bug reports and keep the bug reports that fulfill our criteria (e.g., the bug report should contain input program and reproduce steps), a systematic manual classification and labeling process is performed. Then, to obtain the bug reports which contain compilable input program, we leverage the LLM with few-shot-learning to extract input programs from bug reports and compile them under JVM. The reasons for filtering the compilable input programs are two folds: first, \eclipse refactoring can only be performed on the input programs contain no syntax errors, however, some input programs in the historical bug reports could be incomplete or not syntax error free. Second, those input programs are served as the seeds for the following up mutations, uncompilable seeds could result in variants contains syntax errors, which could result in invalid input programs for \eclipse. After getting the historical bug reports containing compilable input programs as the seeds. We further ask the LLM to extract refactoring information, like refactoring type, input program, and symptom, from those seed bug reports using few-shot-learning by feeding reports' content. Those information will be used in the following up step. Meanwhile, we also ask the LLM to extract the input program structure template, which is the template to represent the structure of the input program. The purposes to extract input program template structure are following: (1) input program structure template keeps the historical bug-triggering input program's structure, meanwhile, it is abstracted by removing the detailed code logic related with current refactoring, thus making a larger search space while mutating the program. For example, the extracted template for input program on the top is listed in the middle of Figure~\ref{inputprogramexample}. Compared with the input program, the code logic inside the method ``innerMethod()'' is removed, so, when mutating the input program, the code mutation will not be limited on existing code logic. (2) Our extracted templates could benefit template-based testing techniques like \astgen since it could serve as the basis or reference. After the refactoring information extraction, we apply the refactoring-preserving transformation (Definition~\ref{def:rptdefinition}) by incorporating the bug-triggering input program characteristics (Table~\ref{characteristics}) together with the templates extracted, thus generating variants that both contain more diverse context and 
 are error-prone. For instance, according to existing study~\cite{wang2024empirical}, input programs with Java anonymous class are more likely to trigger refactoring engine bugs. By constructing chain-of-thought prompts, we instruct LLM to generate refactoring-preserving variants by incorporating the anonymous class with the extracted template. The program in the bottom of Figure~\ref{inputprogramexample} shows one of the refactoring-preserving variants generated by our tool that successfully reveal one new bug in the last version (2024-09) of \eclipse based on the seed program in the top. For this variant, \eclipse fails to resolve the method in an anonymous class when performing the pull up refactoring, producing an refactored program contains syntax error. We have submitted this bug to the \eclipse developers~\cite{issue1766}.
\section{Methodology}
\label{sec:methodology}

\subsection{Overview}

\begin{figure*}
\centering
\includegraphics[width=0.9\linewidth]{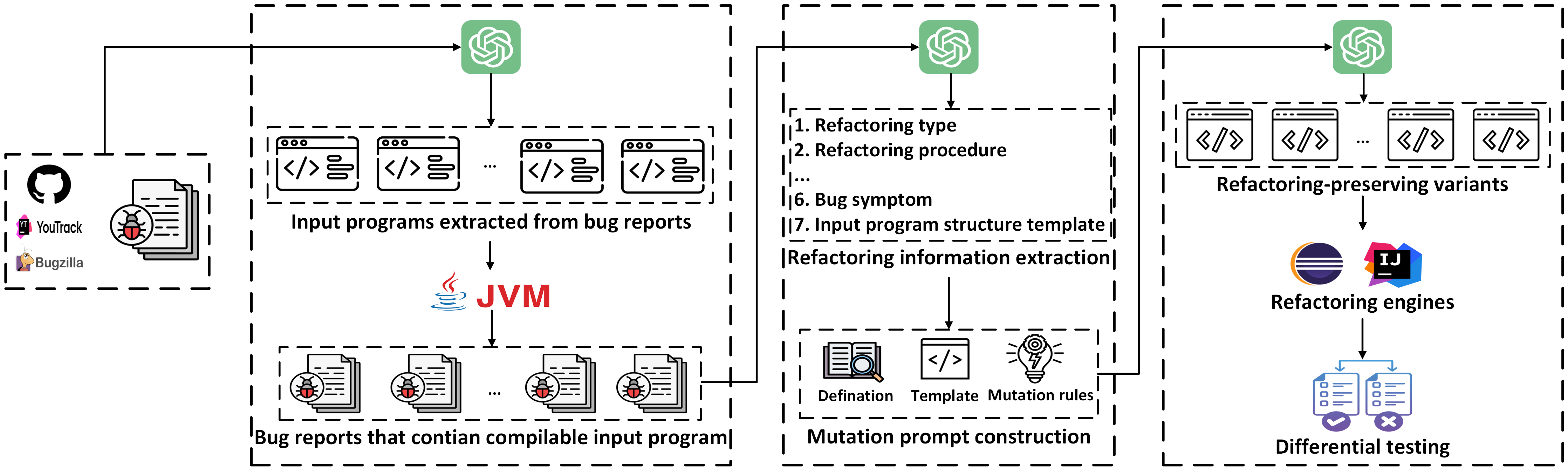}
\caption{Overall workflow of \toolname.}
\centering
\label{workflow}
\end{figure*}


Figure~\ref{workflow} provides a overview of \toolname. Initially, we construct our dataset by gathering historical bug reports from the issue tracker systems of \eclipse and \idea, specifically targeting refactoring engine issues. To ensure relevance, we manually label and classify each report, discarding those unrelated to bugs, such as feature requests. We then employ a Large Language Model (LLM) with few-shot-learning to extract and compile input programs from these reports using a JVM compiler. Only reports containing compilable input programs are retained. Next, the LLM extracts crucial refactoring information from these reports, such as the type of refactoring, procedural details, and input program structure templates. Based on this information, we create mutation prompts reflecting the extracted refactoring details together with the error-prone input program characteristics, as detailed in Table~\ref{characteristics}. These prompts are fed into the LLM to generate diverse mutation variants that preserve the applicability of the original refactoring on the seed input program. Each variant undergoes the same refactoring process in the refactoring engine being tested, and we employ differential testing to determine whether these variants expose any bugs.

\subsection{Dataset Construction}

\subsubsection{Mining historical issues} This step aims to collect a broad range of representative refactoring engine bugs. In this study, we target the two most popular refactoring engines as subjects, including (1) \jdt from \eclipse, and (2) the Java refactoring component of \idea. There are diverse categories of bug reports with different purposes, such as feature requests and questions. Hence, we need to identify the bug reports related with refactoring engine bugs only. Specifically, following existing studies~\cite{sun2016toward,shen2021comprehensive}, we collect bug reports whose title or discussions contain at least one refactoring bug-relevant keyword (i.e., ``refactoring'' and ``refactor'') before the time of our study (July 2024). We focus on bugs that are fixed and not duplicated. Specifically, we consider a bug is fixed if its ``Resolution'' field is set to ``FIXED'' and the ``Status'' field is set to ``RESOLVED'', ``VERIFIED'' or ``CLOSED'' for the \eclipse bug in Bugzilla. After Apr 2022, \eclipse started to migrate their issue trackers to GitHub. To get a complete list of bug reports, we also crawled issues from their GitHub repositories using the GitHub APIs~\cite{githubapis}. For each repository, we search for the fixed issues with the same keywords. For \idea, we collect fixed bugs from its issue tracker using the same keywords. For the issues marked as duplicates by the refactoring engine developers, we will not include them.

\subsubsection{Classification and Labeling Process}
As it is too time-consuming to manually analyze all bugs, we only kept the top 1000 crawled results sorted by fix time for each of our studied refactoring engine. To remove irrelevant issues with refactoring engine bug (e.g., feature requests). Two annotators independently labeled these bug reports. During the labeling process, we filter bug reports according to the following criteria: (1) the issue should be related with refactoring engine, (2) the issue should contain input program, (3) the steps to reproduce bug should be well-illustrated, (4) the bug symptom should be clear (e.g., behavior change). Besides, we also included the bugs newly revealed in~\cite{wang2024empirical}. Following existing approaches~\cite{garcia2020comprehensive,islam2019comprehensive,wang2023compatibility,win2023towards,shen2021comprehensive}, we measured the inter-rater agreement among the annotators via Cohen’s Kappa coefficient. Particularly, the Cohen’s Kappa coefficient was nearly 70\% for the first 10\% bug reports labeling results, thus we conducted a training session about labeling. After that, two annotators
labeled 20\% of bug reports (including the previous 10\%), and Cohen’s Kappa coefficient reached 93\%. After further discussion of the disagreements, Cohen’s Kappa coefficient was always more than 90\% in subsequent labeling iterations (i.e., labeling 20\% \textasciitilde{} 100\% of bug reports with an interval of 10\%). In each labeling iteration, two annotators discussed their disagreements until they reach a consensus. Finally, all bugs were labeled consistently. In total, we obtained \issueNumberofEclipse{} and \issueNumberofIDEA{} bug reports from \eclipse and \idea, respectively.

\subsection{Seed Bug Reports Selection}

\begin{table}
\centering
\caption{The prompt template used to extract input program from historical bug reports.}
\label{extractinputprogram}
\begin{adjustbox}{width=0.49\textwidth, center}
\begin{tabular}{l} 
\toprule
\begin{tabular}[c]{@{}l@{}}You are a software testing expert. I will give you some historical refactoring \\engines bug reports in the following conversations, you need to extract the \\input programs together with their corresponding class names from the bug \\report. The extracted information should be in JSON format, you should \\only return me the extracted input program in the json format, not any \\natural language. I will give you examples following:~\textbf{\{Example\}}\end{tabular}  \\
\bottomrule
\end{tabular}
\end{adjustbox}
\end{table}

The purpose of this step is to select the historical bug reports that contain high-quality seed input programs since they are vital for mutation-based testing~\cite{chen2017learning,zhao2022history,gao2024selecting,li2024boosting}. Specifically, to generate and grow a set of diverse input programs, the input programs should be complete and contain no syntax errors because it serves as the seed for subsequent mutations. Since it is too time-consuming and labor-intensive to validate input program in each bug report manually, so we adopt an automatic way by using a combination of few-shot-learning of LLM and JVM compiler. Different from traditional natural language corpus (e.g., news), a historical bug report (e.g., a bug report for Eclipse~\cite{pullUpRefactoringIssue1533}) is a mixture of several elements such as input program, reproduction procedure (e.g., parameters for invoking a refactoring), environment description, and etc. Our goal is to extract input programs from the bug reports of refactoring engines. 
The input programs of bug reports are challenging to be extracted automatically because the input programs are usually embedded in a bug report within a webpage together with the natural language descriptions without any specific HTML tags or fields that make the relevant code snippets representing the input programs (e.g., the bug reports from Bugzilla are included as plain text comments~\cite{bugzillaextractexample}). Some input programs contain code comments with English texts, further make it difficult to distinguish them from other elements. 

To tackle the aforementioned challenge, we leverage the power of LLM with few-shot learning~\cite{wang2020generalizing,deng2024large} from a small number of manual examples to extract bug-triggering input programs from the large amount of historical bug reports in our dataset. Specifically, the prompt to extract input program from bug report is in Table~\ref{extractinputprogram}, we manually constructed two examples in order to instruct the LLM. For each bug report, we create a new conversation thread in LLM during extraction. To construct the golden standard of the evaluation, we randomly sampled 100 historical bug reports from our dataset obtained through previous step, then we manually extracted input programs from those bug reports to serve as ground truth. To evaluate the effectiveness of our proposed method, we inspected the extraction results for those 100 historical bug reports against ground truth. In total, 98 out of 100 historical bug reports' input programs are correctly extracted, indicating relatively high accuracy in the input program extraction. Only two input programs have been incorrectly extracted because the input programs are incomplete code snippets only contain few lines of code and deeply interleaved with the natural language descriptions. After extracting input programs from collected bug reports, we compiled them with Oracle JDK 22.0.1 using the ``javac'' command. As messages from compilers are automatically generated, they follow strict formats. Upon completion of compilation, we parse the compilation messages to determine whether the JVM compiler accepts or rejects a test program. We filter out the bug reports whose input programs that are uncompilable. Finally, \uncompilableIssueNumber{} out of \totalIssueNumber{} bug reports are filtered out, leaving \compilableIssueNumber{} bug reports remained as seeds, the detailed information is in Table~\ref{ourseeds}. We calculated the mean and median lines of code for the compilable input programs derived from both \eclipse and \idea, finding that the mean and median values were both 11 and 9 lines, respectively.

\begin{table}
\centering
\caption{The seed bug reports information.}
\label{ourseeds}
\begin{adjustbox}{width=0.48\textwidth, center}
\begin{tabular}{ccccc} 
\toprule
\textbf{Source}         & \textbf{Initial} & \textbf{Compilable} & \textbf{Mean LOC} & \textbf{Median LOC}  \\ 
\hline
\eclipse & 245              & 101                 & 11                & 9                    \\
\idea    & 213              & 66                  & 11                & 9                    \\ 
\hline
Total                   & 458              & 167                 & --                 & --                    \\
\bottomrule
\end{tabular}
\end{adjustbox}
\begin{tablenotes}
\footnotesize
\item{
Initial = Number of initial bug reports, Compilable = Number of bug reports that contain compilable input program, Mean LOC = Mean of lines of code for the compilable input programs, Median LOC = Median of lines of code for the compilable input programs.
}
\end{tablenotes}
\end{table}

\subsection{Refactoring Information Extraction}

For the seed historical bug reports obtained through the previous steps, we continue to extract the refactoring information from the bug reports by few-shot-learning. The purpose of this step is to get the necessary information (i.e., refactoring type and input program structure template) for the subsequent step. Meanwhile, by parsing the bug reports, we let LLM acts in a chain-of-thought way. Table~\ref{extractprompt} shows the prompt used to extract refactoring information from historical bug reports. During this step, we aim to get the input program structure template of the input program. This includes the critical program elements and structure related with the applied refactoring in the input program which are essential to trigger bug. Based on the template, we can generate different variants with enriched context and the refactoring applied to the input program is still applicable for the mutated variants. We adopt one-shot-learning for this step. Specifically, we manually construct one example, and then we incorporate the example into the prompt so as to instruct LLM to extract critical information for the other bug reports. The middle of Figure~\ref{inputprogramexample} shows the extracted template for the top input program.

\begin{table}
\centering
\caption{The prompt used to extract refactoring information from historical bug reports.}
\label{extractprompt}
\begin{tabular}{l} 
\toprule
\begin{tabular}[c]{@{}l@{}}You are a software testing expert. I will give you some historical bug \\reports for the refactoring engines. You need to extract the following \\information from the bug reports:\\\textbf{1}. Refactoring type;\\\textbf{2}. Input programs;\\\textbf{3}. Refactored programs;\\\textbf{4}. Refactored program elements information (including element name, \\type, and positions.);\\\textbf{5}. Refactoring procedures;\\\textbf{6}. Bug symptoms;\\\textbf{7}. Input program structure template. \\The extracted information format should be \textbf{\{Format\}}.\\The following is one example: \textbf{\{Example\}}\end{tabular}  \\
\bottomrule
\end{tabular}
\end{table}

\subsection{Mutation Strategy}

For each template extracted above, we construct mutation strategy according to the bug-triggering input program characteristics obtained from previous study~\cite{wang2024empirical}. Input programs contain specific language features (e.g., lambda) or having complex class structures are more likely to trigger refactoring engine bugs and the top error-prone bug-triggering input program characteristics are related with Java language features (e.g., lambda expression, and Java generics), and complex class relationships (e.g., inner class, and anonymous class)~\cite{wang2024empirical}, so, we choose three most error-prone input program characteristics as shown in Table~\ref{characteristics} in our study. Although there are diverse individual input program characteristic (38 types in~\cite{wang2024empirical}) that are error-prone, and different individual characteristic can also be combined to generate more complex input programs that could potentially trigger more bugs in the refactoring engine, we only select three of most error-prone characteristics to set a bound for the large search space of possible combinations. 

Our prompt template used to perform the mutation is shown in Table~\ref{mutationprompt}. Our prompt design is inspired by chain-of-thought (CoT) prompting, where instead of directly generating the final output, the prompt asks the model to finish task by breaking down the problem into sequential steps. Specifically, we first let LLM to understand current refactoring by giving the refactoring definition. Then, we give the LLM the input program structure template obtained from previous step and the bug-triggering input program characteristic in Table~\ref{characteristics}, and ask it to generate variants while preserving the original refactoring. Next, we add some extra requirements in our prompt template. For example, to reduce the uncompilable variants, we require the generated variants should conformance with specific JDK version. The variants are obtained through the Refactoring-preserving Transformation (RPT) as shown in Definition~\ref{def:rptdefinition}, which means the same refactoring should be applicable on both the original input program and its variant.
\noindent \begin{definition}[Refactoring-Preserving Transformation (RPT)]
\label{def:rptdefinition}
Let $P_1$ be an input program and $E_1$ a program element within $P_1$ targeted for refactoring. Consider a refactoring operation $O$ that is applicable to $E_1$. A transformation function $\text{Trans()}$ defines a transformation such that $P_2 = \text{Trans}(P1)$, where $P_2$ is the transformed program. This transformation is deemed refactoring-preserving if there exists a program element $E_2$ in $P_2$ and the refactoring operation $O$ remains applicable to $E_2$. Thus, the transformation $\text{Trans()}$ preserves the applicability of the refactoring operation from $P_1$ to $P_2$.
\end{definition}

Figure~\ref{inputprogramexample} bottom shows one variant generated by the LLM using the prompt template in Table~\ref{mutationprompt} whose bug-triggering characteristic is set to Java anonymous class. This variant is refactoring-preserving, which means the ``Pull Up'' refactoring could still be applied on the ``methodToBePulledUp()'' method on variant. This variant successfully triggered one bug in the newest version of \eclipse (2024-09), resulting in an uncompilable refactored program. We have reported the bug together with the variant input program and bug reproduce steps to the \eclipse's issue tracker systems~\cite{issue1766}. We measure the LLM's ability to perform RPT in RQ1.

\begin{table}
\centering
\caption{Input program characteristics applied for refactoring-preserving transformation.}
\label{characteristics}
\begin{tabular}{cl} 
\hline
Characteristic  & \multicolumn{1}{c}{Description}                                                                                                                    \\ 
\hline\hline
Lambda          & \begin{tabular}[c]{@{}l@{}}Anonymous functions used to implement functional \\interfaces with a more streamlined syntax\end{tabular}               \\
Java generics   & \begin{tabular}[c]{@{}l@{}}Java generics allow to create classes, interfaces, \\and methods that operate with unspecified types\end{tabular}       \\
Anonymous class & \begin{tabular}[c]{@{}l@{}}Class defined without a name, often used for one-time \\implementations of interfaces or abstract classes\end{tabular}  \\
\hline
\end{tabular}
\end{table}

\begin{table}
\centering
\caption{The prompt template used to perform mutations.}
\label{mutationprompt}
\begin{tabular}{l} 
\toprule
\begin{tabular}[c]{@{}l@{}}Now, I will give the definition of the current refactoring, you need to \\understand it. You need to make sure the original refactoring could \\still be applied on the variant.\\\textbf{1}. \textbf{\{Refactoring Type\}}: \textbf{\{Definition\}}\\\textbf{2}. To expose more bugs in the refactoring engines, please generate \\edge case variant considering the \textbf{\{Characteristic\}} in current \\refactoring scenario. You need to generate the variant according \\to the Input Program Structure Template, it is \textbf{\{Template\}}.\\\textbf{3}. You should give me the variant, the program elements \\to be refactored, and the procedures to refactoring.\\\textbf{4}. The generated variant should not contain any syntax errors. \\The Java program you generated should conformance with the \\JDK \textbf{\{Version\}} standard.\\Please generate one edge case variant considering different edge \\usage scenarios of \textbf{\{Characteristic\}} based on the template. The~\\variant format should be \textbf{\{Format\}}.\end{tabular}  \\
\bottomrule
\end{tabular}
\end{table}

\subsection{Differential Testing}
We manually refactor the variants in the newest version of refactoring engines by applying the same refactoring as in the seed bug reports, and compare the refactoring results of \eclipse and \idea using differential testing. To ensure the reliability and objectivity of the manual process, two annotators independently refactor each variant in the IDEs according to the reproduce procedures in the seed bug reports. Then, two annotators hold a meeting to resolve their disagreements. Based on prior work on testing refactoring engines~\cite{daniel2007automated}, we adopt the following oracles:


\noindent \textbf{Uncompilable Oracle.} This checks if any of the refactored program contains syntax errors.

\noindent \textbf{Warning Status Oracle.} This compares and checks if the warning status from different refactoring engines are different. For example, one refactoring engine may produce a warning message but the other does not (this might occur due to the overly weak or strong preconditions checking in different refactoring engines).

\noindent \textbf{Differential Oracle.} This checks if (1) the refactoring has been performed successfully, (2) the refactored programs contain no syntax error, and (3) whether the refactored programs are the same.

As each oracle violation may indicate potential bugs in refactoring engines, we manually verify all violations. 
Before submitting the issues, we search in the corresponding bug-tracking systems of the target IDEs to avoid submitting duplicate reports. The search process involves looking for the keywords representing the refactoring type and bug symptoms. If any bug reports with similar input program and same symptom are found, we consider the bugs were already reported and will not report them.

\section{Evaluation}
\label{sec:experiments}

\subsection{Implementation}
\label{sec:implementation}

As shown in Table~\ref{seeds}, we randomly select five historical bug reports from our seed dataset for our experiment. Specifically, two seeds are from \eclipse and three are obtained from \idea. These seeds cover five refactoring types and two symptoms. We use OpenAI API~\cite{openaiapi} to programmatically invoke the ChatGPT. The model type is gpt-4o-mini with the default settings~\cite{defaultsetting} and the knowledge cut-off date is Oct 2023. For each seed, we iterate the characteristics listed in Table~\ref{characteristics}, and for each characteristic, we ask the LLM to generate ten variants. In conclusion, we generate 30 variants for each seed considering three different characteristics. In total, 150 variants are generated for the five seeds in Table~\ref{seeds}. We set the JDK version in Table~\ref{mutationprompt} to 22.0.1, and the corresponding refactoring definition is obtained from~\cite{fowler2018refactoring}. We manually tested the input program variants generated by our tool in the latest version of \eclipse (2024-09) and \idea (2024.2.4). All experiments are run on a workstation with 2.6GHz 6-core Intel Core i7 CPU and Windows 10, 64-bit operating system.

\begin{table}
\centering
\caption{Seed historical bug reports used in our experiment.}
\label{seeds}
\begin{tabular}{ccccc} 
\toprule
\textbf{ID} & \textbf{Source} & \textbf{Issue No.} & \textbf{Refactoring Type} & \textbf{Symptom}  \\ 
\hline\hline
S-1         & Eclipse         & 1533               & Pull up                   & Compile error     \\
S-2         & Eclipse         & 1529               & Inline method             & Compile error     \\
S-3         & IDEA            & 142361             & Extract variable          & Compile error     \\
S-4         & IDEA            & 354116             & Make static               & Behavior change   \\
S-5         & IDEA            & 354122             & Extract method            & Compile error     \\
\bottomrule
\end{tabular}
\end{table}

\subsection{Research Questions}
We investigate the following research questions in our experiments:

\noindent \textbf{RQ1:} \emph{How effective is \toolname in detecting refactoring engine bugs?}

\noindent \textbf{RQ2:} \emph{How does \toolname perform compared to 
 other baseline?}

\noindent \textbf{RQ3:} \emph{Which input program characteristics are more useful to reveal bugs?}

\noindent \textbf{RQ4:} \emph{What is the contribution of input program structure template?}

\section{Evaluation Results}
\label{sec:result}

\subsection{RQ1: Effectiveness}

\begin{table*}
\centering
\caption{Experimental result of \toolname.}
\label{initial_result}
\begin{adjustbox}{width=0.8\textwidth, center}
\begin{tabular}{c|c|c|c|c|c|c|ccc|cc} 
\toprule
\multirow{2}{*}{\textbf{Refactoring}} & \multirow{2}{*}{\textbf{Template}} & \multirow{2}{*}{\textbf{ET (s)}} & \multirow{2}{*}{\textbf{TGV}} & \multirow{2}{*}{\textbf{MT (s)}} & \multirow{2}{*}{\textbf{CV}} & \multirow{2}{*}{\textbf{RPV}} & \multicolumn{3}{c|}{\textbf{Oracles}}      & \multicolumn{2}{c}{\textbf{Bugs}}  \\ 
\cline{8-12}
                                      &                                    &                                  &                               &                                  &                              &                              & \textbf{UC} & \textbf{WS} & \textbf{Diff.} & \textbf{EC} & \textbf{IDEA}        \\ 
\hline\hline
\multirow{2}{*}{Extract method}       & Y                                  & 6                                & 30                            & 87                               & 27                           & 27                           & 1           & 0           & 0              & 1           & 0                    \\
                                      & N                                  & 7                                & 30                            & 131                              & 28                           & 28                           & 0           & 0           & 0              & 0           & 0                    \\\hline
\multirow{2}{*}{Inline method}        & Y                                  & 8                                & 30                            & 91                               & 26                           & 26                           & 5           & 0           & 0              & 5           & 0                    \\
                                      & N                                  & 7                                & 30                            & 81                               & 20                           & 20                           & 3           & 0           & 0              & 3           & 0                    \\\hline
\multirow{2}{*}{Extract variable}     & Y                                  & 6                                & 30                            & 78                               & 30                           & 30                           & 0           & 0           & 0              & 0           & 0                    \\
                                      & N                                  & 6                                & 30                            & 73                               & 25                           & 25                           & 0           & 0           & 0              & 0           & 0                    \\\hline
\multirow{2}{*}{Pull up}              & Y                                  & 7                                & 30                            & 105                              & 20                           & 20                           & 8           & 1           & 0              & 7           & 2                    \\
                                      & N                                  & 8                                & 30                            & 109                              & 20                           & 20                           & 2           & 0           & 0              & 1           & 1                    \\\hline
\multirow{2}{*}{Make static}          & Y                                  & 10                               & 30                            & 104                              & 25                           & 22                           & 0           & 0           & 2              & 0           & 0                    \\
                                      & N                                  & 11                               & 30                            & 179                              & 26                           & 21                           & 1           & 0           & 0              & 1           & 0                    \\ 
\hline\hline
\textbf{Average}                      & --                                  & 7.6                              & 30                            & 103.8                            & 25                           & 24                           & \multicolumn{3}{c|}{--}                     & \multicolumn{2}{c}{--}              \\ 
\hline
\textbf{Total}                        & --                                  & 76                               & 300                           & 1038                             & 247                          & 239                          & 20          & 1           & 2              & 18 (15)     & 3                    \\
\bottomrule
\end{tabular}
\end{adjustbox}
\begin{tablenotes}
\footnotesize
\item{
Template = Whether input program template is used during mutation, ET = Time taken in seconds to extract refactoring information, TGV = Total generated variants, MT = Mutation time for TGV in seconds, CV = Compilable variants, RPV = Refactoring-preserving variants; Oracles: UC = Uncompilable Oracle, WS = Warning Status Oracle, Diff. = Differential Oracle; EC = \eclipse, IDEA = \idea.
}
\end{tablenotes}
\end{table*}

Table~\ref{initial_result} presents detailed experimental results obtained using our tool, \toolname. The ``Refactoring'' column lists the types of refactoring operations tested with our tool. It is important to note that the capabilities of our tool are not confined to the refactoring operations tested; its design is based on historical bug reports, facilitating easy extension to additional refactoring types by incorporating a diverse range of bug reports. Column ``Template'' tells about whether input program template is leveraged during mutation. The third column, ``ET'' indicates the time required to extract refactoring information for each type, which typically takes less than ten seconds. As discussed in ~\ref{sec:implementation}, we generate 30 input program variants for each refactoring type. The ``MT'' column details the time taken to mutate these variants. Given that our tool leverages a large language model (LLM), it inherently generates variants that may contain syntax errors~\cite{zhong2024can,wang2023review}. However, the refactoring engine in \eclipse requires syntax-error-free input programs, otherwise refactoring operations cannot be performed, we subsequently filter the uncompilable variants using a JVM compiler. The ``CV'' column reports the number of compilable variants, with ``Extract variable'' having the highest count due to the simpler nature of its input programs, which facilitates easier mutation. Column ``RPV'' shows the number of refactoring-preserving variants, while most of the compilable variants preserve the original refactoring, there exists variants for ``Make static'' refactoring that are not refactoring-preserved. This happens because the method to be refactored in the input program is mutated to a static method in variants, however, one of the preconditions to perform ``Make static'' refactoring is that the method to be refactored should not already been static~\cite{makestaticrefactoringpreconditions}, thus leading to eight variants cannot applying ``Make static'' refactoring. The ``Oracles'' column calculates the number of variants for each oracle as outlined in our methodology. Predominantly, most bugs are identified by our ``Uncompilable Oracle,'' indicating that the refactoring engine often produces refactored programs with syntax errors. This aligns with previous studies on refactoring engine bugs where compilation errors are the most common symptom~\cite{wang2024empirical}. For the same input program, \eclipse and \idea may yield different syntax-error-free refactored programs due to the different default refactoring configurations employed by each engine. For example, when performing ``Make static'' refactoring for a method having parameters, \idea would declare the parameters as ``final'' during refactoring while \eclipse would not. We manually analyzed the differential results for ``Make static'' refactoring and confirmed that all two variants were correctly refactored. Those two differential results were produced because the default configurations for ``Make static'' refactoring are slightly different in those two refactoring engines as described above, thus leading to two false positive. However, the default configuration for other refactoring operations remain the same. According to previous study~\cite{wang2024empirical}, most of the bugs (97\%) could be triggered by the default initial configuration of refactoring engines, and the default configuration for different refactoring engine could be changed to the same by setting up the same configuration parameters. The last two columns record the number of bugs we identified in \eclipse and \idea, which is \eclipseIssueNumber{} and \ideaIssueNumber{}, respectively. For \eclipse, there are three overlap bugs when testing ``Inline method'' refactoring with and without input program template during mutation stage, thus leading to \eclipsenotoverlapIssueNumber{} unique new bugs. Prior to report, we searched the refactoring engine's bug-tracking system to avoid duplicates. Notably, ``Pull up'' refactoring accounts for \pullupIssueNumber{} out of the \foundIssueNumber{} reported bugs, primarily due to its involvement with complex class relationships that complicate the refactoring process.

\begin{table}
\centering
\caption{Detailed information for our submitted bug reports.}
\label{tab:submittedbugreports}
\begin{adjustbox}{width=0.5\textwidth, center}
\begin{tabular}{cccccc} 
\toprule
\textbf{ID} & \textbf{IDE} & \textbf{Issue No.} & \textbf{Refactoring Type} & \textbf{Symptom}   & \textbf{Status}              \\ 
\hline\hline
B-1         & Eclipse      & 1785               & Extract Method            & Compile error      & Submitted                    \\
B-2         & Eclipse      & 1824               & Make Static               & Compile error      & \textbf{\textbf{Confirmed}}  \\
B-3         & Eclipse      & 1783               & Inline Method             & Compile error      & Submitted                    \\
B-4         & Eclipse      & 1781               & Inline Method             & Compile error      & Submitted                    \\
B-5         & Eclipse      & 1780               & Inline Method             & Compile error      & \textbf{\textbf{Fixed}}      \\
B-6         & Eclipse      & 1779               & Inline Method             & Compile error      & Submitted                    \\
B-7         & Eclipse      & 1778               & Inline Method             & Compile error      & Submitted                    \\
B-8         & Eclipse      & 1777               & Pull Up                   & Compile error      & Submitted                    \\
B-9         & Eclipse      & 1776               & Pull Up                   & Compile error      & Submitted                    \\
B-10        & Eclipse      & 1775               & Pull Up                   & Compile error      & Submitted                    \\
B-11        & Eclipse      & 1774               & Pull Up                   & Failed refactoring & Submitted                    \\
B-12        & Eclipse      & 1773               & Pull Up                   & Compile error      & \textbf{Fixed}               \\
B-13        & Eclipse      & 1772               & Pull Up                   & Compile error      & Submitted                    \\
B-14        & Eclipse      & 1766               & Pull Up                   & Compile error      & Submitted                    \\
B-15        & Eclipse      & 1823               & Pull Up                   & Compile error      & \textbf{\textbf{Fixed}}  \\
B-16        & IDEA         & 364110             & Pull Members Up           & Compile error      & \textbf{\textbf{Confirmed}}  \\
B-17        & IDEA         & 362805             & Pull Members Up           & Compile error      & \textbf{Confirmed}           \\
B-18        & IDEA         & 362804             & Pull Members Up           & Compile error      & \textbf{Confirmed}           \\
\bottomrule
\end{tabular}
\end{adjustbox}
\begin{tablenotes}
\footnotesize
\item{
The issues of \idea, and \eclipse can be found at https://youtrack.jetbrains.com/issue/IDEA-XXX, and https://github.com/eclipse-jdt/eclipse.jdt.ui/issues/XXX, where ``XXX'' can be replaced with the concrete numbers in \textbf{Issue No.}.
}
\end{tablenotes}
\end{table}

Table~\ref{tab:submittedbugreports} presents comprehensive details of the \foundIssueNumber{} bug reports we submitted, including \withTemplateIssueNumber{} bugs identified in RQ1 and \withoutTemplateNotOverlapIssueNumber{} discovered in RQ4. The ``IDE'' column identifies the refactoring tool under evaluation. The columns ``Issue No.'' and ``Refactoring Type'' specify the respective issue numbers in the bug-tracking systems and their corresponding refactoring types. The ``Symptom'' column provides a summary of the symptoms associated with each bug, while the final column reports the current status of each issue. As of the submission of this paper, \totalconfirmedIssueNumber{} have been confirmed by developers. Notably, ``Pull Up'' and ``Inline Method'' refactoring revealed the highest number of bugs, with eleven and five instances respectively, predominantly resulting in compile errors. This observation supports findings from existing studies~\cite{wang2024empirical}, which indicate that compile errors are the most prevalent symptom in refactoring engine bugs. Furthermore, four out of the eighteen bugs have been officially confirmed by their respective developers, highlighting the ongoing engagement between our research efforts and the development community to enhance the reliability of refactoring tools.

\subsection{RQ2: Comparison with Baseline}

We selected the state-of-the-art Gligoric et al.~\cite{gligoric2013systematic} testing approach as our baseline, which perform systematic testing of refactoring engines by applying refactoring operations at randomly chosen program elements within real-world projects. To ensure a fair comparison, we replicate this process by employing a large language model (LLM) to propose refactoring for each input program derived from the seed bug reports. Specifically, for each program, the LLM is instructed to suggest ten possible refactoring operations, mirroring the number of transformations we have predetermined for each category, such as ``Lambda''. Subsequently, these suggested refactoring operations are implemented in \eclipse and \idea to uncover potential bugs.

\begin{table}
\centering
\caption{Different transformations in our tool compared to the baseline.}
\label{distributionofbugsforcharacter}
\begin{adjustbox}{width=0.48\textwidth, center}
\begin{tabular}{c|c|c|c|c|c|cc} 
\toprule
\multirow{2}{*}{\textbf{Transformation}}                       & \multirow{2}{*}{\textbf{ET (s)}} & \multirow{2}{*}{\textbf{TGV}} & \multirow{2}{*}{\textbf{MT (s)}} & \multirow{2}{*}{\textbf{\textbf{CV}}} & \multirow{2}{*}{\textbf{RPV}} & \multicolumn{2}{c}{\textbf{Bugs}}              \\ 
\cline{7-8}
                                                               &                                  &                               &                                  &                                       &                               & \textbf{\textbf{EC}} & \textbf{\textbf{IDEA}}  \\ 
\hline
$\toolname_L$                                                  & 8                                & 50                            & 146                              & 46                                    & 45                            & 4                    & 2                       \\
$\toolname_G$                                                  & 8                                & 50                            & 91                               & 39                                    & 37                            & 3                    & 0                       \\
$\toolname_A$                                                  & 9                                & 50                            & 131                              & 43                                    & 43                            & 6                    & 0                       \\
Gligoric et al.~\cite{gligoric2013systematic} & 9                                & 50                            & 152                              & 45                                    & 45                            & 0                    & 0                       \\
\bottomrule
\end{tabular}
\end{adjustbox}
\begin{tablenotes}
\footnotesize
\item{
ET = Extract refactoring information time, time is in seconds; TGV = Total generated variants, MT = Mutation time for TGV in seconds, CV = Compilable variants, RPV = Refactoring-preserving variants, EC = \eclipse, IDEA = \idea.
}
\end{tablenotes}
\end{table}

Table~\ref{distributionofbugsforcharacter} provides comparative results for each component of our tool versus the baseline. In the notation used, $\toolname_L$ denotes the application of only the Lambda transformation, as detailed in Table~\ref{characteristics}. Similarly, $\toolname_G$ and $\toolname_A$ indicate exclusive application of Java generics and anonymous class transformations, respectively, within our tool. Based on the five seeds in Table~\ref{seeds}, we generate ten variants for each seed by implementing each transformation, resulting in a total of 50 variants. The concluding two columns display the number of bugs identified in \eclipse and \idea by each component of our tool and the baseline, respectively. Notably, our tool successfully detected \withTemplateIssueNumber{} new bugs. Without performing mutation on the input program, Gligoric's approach fails to find any bugs using the same set of seed input programs as our approach. This substantial discrepancy highlights the enhanced efficacy of our tool in identifying potential issues within refactoring engines.

\subsection{RQ3: Contribution of Different Characteristics}

Table~\ref{distributionofbugsforcharacter} shows the number of bugs identified by applying various input program characteristics through our tool. Specifically, $\toolname_L$, $\toolname_G$, and $\toolname_A$ represent configurations where only Lambda transformations, Java generics, and anonymous class transformations are applied, respectively. The results show that lambda and anonymous class transformations each exposed six bugs, whereas Java generics revealed three bugs. In terms of mutation time, lambda transformations are the most time-consuming ones, taking 146 seconds. Java generics generated the least number of compilable variants, with only 39 out of 50, largely due to the complex syntax associated with generics which makes it challenging to generate syntactically correct variants. The majority of bugs, \withTemplateEclipseIssueNumber{} out of \withTemplateIssueNumber{}, were detected in \eclipse, with anonymous class transformations revealing six, lambda transformations revealing four, and Java generics uncovering three. This distribution suggests that developers of \eclipse should particularly focus on improving refactoring operations that involve complex class relationships, type inference, and code transformations. In future, we plan to consider further enhancements in the bug detection capability by combining individual characteristics or by generating variants that incorporate a broader range of diverse and bug-triggering input program characteristics~\cite{wang2024empirical}.

\subsection{RQ4: Effectiveness of the Template}

We conducted an ablation study to elucidate the specific contributions of the input program template to the mutation process. For this study, we modified the mutation prompts in Table~\ref{mutationprompt} by substituting the template information with details from the actual input program. Specifically, we revised the prompt ``You need to generate the variant according to the Input Program Structure Template, it is \{Template\}.'' to ``You need to generate the variant according to the Input Program, it is \{Input Program\}.'' Additionally, the instruction ``Please generate one edge case variant considering different edge usage scenarios of \{Characteristic\} based on the template.'' was altered to ``Please generate one edge case variant considering different edge usage scenarios of \{Characteristic\} based on the Input Program.'' The input programs utilized for this process were derived from our refactoring information extraction step. Except for these modifications to the mutation prompts, all other experimental settings remained unchanged.

\begin{figure}
\centering
\includegraphics[width=0.6\linewidth]{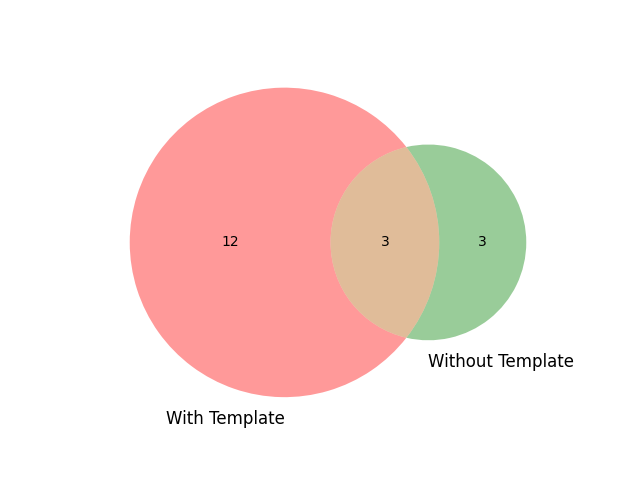}
\caption{The venn diagram for the number of bugs detected by \toolname with and without input program template.}
\centering
\label{templatevenn}
\end{figure}

Table~\ref{initial_result} presents the outcomes of the experiment conducted without utilizing the input program template in the ``Template'' column marked as ``N''. The lst two columns indicate the number of bugs revealed: three for the ``Inline Method'' refactoring, two for the ``Pull Up'' refactoring, and one for the ``Make Static'' refactoring. In comparison to the original results listed in Table~\ref{initial_result}, where \withTemplateIssueNumber{} bugs were detected, only \withoutTemplateIssueNumber{} bugs were identified when the template was not used. This reduction in bug detection can be attributed to the constraints imposed by using specific input programs, which significantly narrows the LLM's search space for generating variants. For instance, in the template shown in Figure~\ref{inputprogramexample} middle, the code within the method methodToBePulledUp() is abstracted to suggest potential logic interacting with the outer class context. Conversely, in the input program depicted at the top of the figure, the actual logic within the method is retained, restricting the LLM to perform further mutations based on this specific method logic, thereby resulting in less variant diversity. We also observe that the approach without prompt template can only detect bugs with uncompilable oracle (UC). Without getting guidance on the abstracted program structures, the approach without template can only rely on knowledge about the detailed code logic and Java syntax to generate variants that result in syntactically incorrect programs, which limits its capability in detecting bugs that require more complex oracles (i.e., Warning Status Oracle and Differential Oracle). We conducted further analysis on the bugs revealed with and without the template to assess overlap and uniqueness. The Venn diagram in Figure~\ref{templatevenn} shows that \overlapIssueNumber{} bugs are overlapped, constituting \withoutTemplateOverlapIssueNumberRatio{} of the bugs detected by \toolname without using the template. Additionally, \withTemplateNotOverlapIssueNumber{} and \withoutTemplateNotOverlapIssueNumber{} unique bugs were identified exclusively with and without the template, respectively, illustrating the significant role of the input program template in facilitating diverse code mutations.

\section{Implication}
\label{sec:implication}
Based on our study and analysis, we discuss the implications for researchers and refactoring engine's developers.

\noindent \textbf{Implication for Researchers.} 
Our study and automated testing framework lay the foundation for future research in three promising directions. First, our seed dataset serve as cornerstone for future research in automated refactoring engine testing. The lines of code (LOC) analysis result in Table~\ref{ourseeds} indicates that bug-revealing input program is usually having a small size, with mean and median LOC is 11 and 9, respectively. \emph{This observation can be leveraged for improving the effectiveness of test generation for refactoring engine validation. A test generation tool should focus on generating small but complex input programs rather than big but simple ones.} Second, bug-triggering input program characteristics are critical for refactoring engine testing. Input programs own specific characteristics (e.g., anonymous class) are more error-prone. This finding indicates that when generating test program to reveal bugs, we should take the program characteristics into account. \emph{Automated testing techniques in the area of compiler, static analyzer, and others which take program as input could all benefit from incorporating the error-prone input program characteristics.} Our study serves as a preliminary study to motivate future research on using a richer set of transformations based on the bug-triggering input program characteristics for improving the reliability of static analysis tools. \emph{Third, the input program structure template obtained through our tool could be used for template-based techniques as their references for designing templates, thus improving effectiveness.}

\noindent \textbf{Implication for Developers.} 
Our study identifies new bugs in both \eclipse and \idea leveraging the historical bug reports and error-prone input program characteristics. \emph{Developers of refactoring engine should consider adding a checklist for those characteristics when testing their refactoring operations in IDEs.} This could be obtained by analyzing the historical bug reports. Since manually writing test input programs for each refactoring could be labor-intensive and time-consuming, \emph{refactoring engine developers could leverage the LLM as assistant while testing considering its effectiveness}. Developers should set a higher priority for input programs that contain certain characteristics (e.g., Lambda), since they tend to have a higher bug-revealing capability. 
\section{Threats to Validity}
\label{sec:threats}
We identify the following threats to the validity:

\noindent \textbf{Internal.} The internal threat to validity mainly lies in our manual classification and labeling of refactoring engine bugs, which may have subjective bias or errors. To reduce this threat, we referred to the previous studies~\cite{shen2021comprehensive,yang2011finding}, and then adopted an open-coding scheme. During the labeling process, two annotators independently labeled bugs, any disagreement was discussed at a meeting until a consensus was reached. As our dataset and our approach only focus on Java input programs generation, the findings may not generalize for other programming languages beyond Java.

\noindent\textbf{External.} The external threat to validity mainly lies in the dataset used in our study. To reduce this threat, we systematically collected refactoring engine bugs as presented in Section~\ref{sec:methodology}. To ensure the diversity and generalization of the considered refactoring engines, we choose two of the most popular refactoring engines (i.e., \eclipse and \idea) as our studied target.
\section{Related Work}
\label{sec:relatedwork}

\noindent \textbf{LLM for testing.}
Recent advancements in Large Language Models (LLMs) have significantly boost software testing methodologies. TitanFuzz introduces LLMs to fuzz deep learning libraries by generating seed inputs specified through API-related prompts and employing mutation strategies like code masking, followed by LLM completion~\cite{deng2023large}. FuzzGPT extends this by using few-shot prompts, providing examples of code and descriptions to generate test cases for deep learning libraries~\cite{deng2024large}. Similarly, LAST leverages LLMs to validate SMT solvers by generating diverse formulas~\cite{sun2023smt}. WhiteFox marks a significant advancement as the first white-box compiler fuzzer that uses source-code information with LLMs to test compiler optimizations, uncovering deep logic bugs in deep learning compilers~\cite{yang2024whitefox}. We are the first to leverage LLM for refactoring engine testing by combing historical bug reports and error-prone input program characteristics.
\section{Conclusion}
\label{sec:conclusion}
In this paper, we introduced a novel approach to test refactoring engine based on historical bug reports and bug-triggering input program characteristics empowered by Large Language Model (LLM). \toolname does not require manually designing templates and can easily generalize to diverse refactoring types. The experimental results show that \toolname can detect various new bugs in both \eclipse and \idea. In total, we identified \foundIssueNumber{} new bugs in the latest version of both refactoring engines. By the submission time of our paper, \totalconfirmedIssueNumber{} bugs were confirmed, \submittedIssueFixedNumber{} of them were fixed.

\textbf{Data Availability.}
The data is available at~\cite{anonymousrepolink}.

\section*{Acknowledgments}
This work is supported by the Natural Sciences and Engineering Research Council of Canada (NSERC) Discovery Grants RGPIN-2024-04301.

\bibliographystyle{IEEEtran}
\bibliography{ref}
\end{document}